\documentclass{llncs}
\usepackage{amssymb}

\title{Preliminaries to an Investigation of \\
       Reduced Product Set Finance}
\author{J.A. Bergstra\inst{1}\fnmsep\inst{2} and C.A. Middelburg\inst{1}}
\institute{Informatics Institute, Faculty of Science,
           University of Amsterdam, \\
           Science Park~904, 1098~XH Amsterdam, the Netherlands \\
           \email{J.A.Bergstra@uva.nl,C.A.Middelburg@uva.nl}
           \and
           Department of Computer Science, School of Physical Sciences,
           Swansea University, \\
           Singleton Park, Swansea SA2~8PP, United Kingdom
          }

\begin{document}
\maketitle

\begin{abstract}
Principles of financial product synthesis from a few basic financial
products constitute an interesting research topic inspired by Islamic
finance.
We make an effort to answer general questions that should be answered
before starting to investigate the main issues concerning this topic
with the formalization of financial products and principles of financial
product synthesis.
We also outline the outcome of some preparatory explorations, which have
been conducted with the purpose to form a reasonable preliminary picture
of the details of financial products that are relevant to the study of
the principles of financial product synthesis.
\par\addvspace{1.5ex} \small {\sl Keywords:}
reduced product set finance, financial product synthesis,
Islamic finance.
\nolinebreak
\end{abstract}

\section{Introduction}
\label{sect-introduction}

The idea of considering only financial products that are synthesized
from a few basic financial products according to certain principles is
both appealing from a scientific point of view and directly relevant to
the practice of finance.
In science, at least in computer science, it is common practice to start
developing a sound understanding of complex concepts by deliberately
simplifying matters like this.
However, the idea is also actually used in the practice of Islamic
finance.
This makes it very attractive to investigate issues such as the issue
whether principles of financial product synthesis can be formulated
that, if applied to products that are legal according to Islamic law,
lead to synthesized financial products that are also legal according to
Islamic law and the dependent issue which financial products that are
not legal according to Islamic law can be approximated by financial
products that are synthesized according to legality-preserving
principles from legal products.

We believe that before such an investigation is conducted, answers
to certain general questions should be given.
Among the most important general questions are the question whether the
investigators can conduct the investigation impartially, the question
whether the investigation can be conducted without assumptions about the
validity of ethical principles, the question whether the investigation
is timely, and the question whether the object of the investigation has
an application perspective.
We believe that, in order to achieve reliable results, the formalization
of financial products and principles of financial product synthesis must
form part of the investigation.
We expect that methods and techniques developed in computer science will
play an important part in the formalization.
An interesting general question raised by this is the question what is
the relevance of the investigation for computer science.

Among the more specific questions related to the formalization of
financial products and principles of financial product synthesis are the
question which details of financial products are needed to formulate
principles of financial product synthesis and the question which details
of financial products are needed to determine their legality.
The more specific questions like these ones cannot be answered fully
before the investigation is conducted.
We have done some preparatory explorations by means of examples with the
purpose to form a reasonable preliminary picture of the details referred
to above.

In this paper, which was written in preparation to the investigation
brought up above, we make an effort to answer the above-mentioned
general questions and we outline the outcome of the preparatory
explorations.

The paper is organized as follows.
First, we give some classical examples of synthesized financial products
and introduce the kind of finance involved \linebreak[2]
(Section~\ref{sect-synthesis}).
After that, we go into the question whether issues related to Islamic
finance can be investigated from a non-Islamic perspective at all
(Section~\ref{sect-impartiality}). \linebreak[2]
Next, we make an effort to answer a number of general questions to which
answers should be given before the investigation proposed in this paper
is conducted (Section~\ref{sect-questions}).
Following this, we pose questions about the classical synthesized
financial product examples given in Section~\ref{sect-synthesis} and the
prevailing positions concerning their legality
(Section~\ref{sect-more-synthesis}).
Then, we look at a loan with interest, including the particular case of
a savings account with interest, with the purpose to get an impression
of what it takes to formalize simple financial products
(Sections~\ref{sect-loan-with-interest}
and~\ref{sect-savings-account-with-interest}).
Thereafter, we carry out a preparatory exploration of the synthesis of
a savings account with interest in the way known as \emph{tawarruq}
(Section~\ref{sect-exploration}).
Finally, we make some concluding remarks
(Section~\ref{sect-conclusions}).

To clarify the title of this paper, we remark that we will use in this
paper the term ``reduced product set finance'' for finance where only
financial products that are synthesized from a few basic financial
products according to certain principles are considered.

\section{Synthesized Financial Products}
\label{sect-synthesis}

In the financial literature, there are publications which go into
financial products that are synthesized from a few basic financial
products according to certain principles.
The principles in question are compositional methods for synthesizing
financial products from other financial products.
In this section, we give some classical examples of synthesized
financial products and introduce the kind of finance involved.

\subsection{Examples of Synthesized Financial Products}
\label{subsect-examples-synthesis}

From Mahmoud A. El-Gamal~\cite{ElG08a}, we take the following quotation:
\begin{quotation}
\noindent
Now, let's assume that the regulators or religious authorities deemed
each of transactions B and C individually to have more benefit than
harm.
Therefore, there were no prohibitions against either B or C.
Assume finally that A equals B plus C.
Then, we have a legal arbitrage opportunity that can be exploited by
financially engineering A from B and C, knowing that A must be desirable
to have been forbidden in the first place.

To consider a concrete example, let A, the forbidden contract, be
interest-based loans, and let B and C represent spot sales and credit
sales of any goods.
An ancient contract called in Islamic jurisprudence
\emph{bay'u al-'ina}, or same item sale-repurchase, may thus be utilized
to synthesize A from B and C.
Assume that A would result in X lending one hundred dollars to Y, and
receiving one hundred and ten dollars in one year.
Assume that Y owns an asset S, and let B and C represent credit and spot
trades in S.
Now, all Y has to do is to sell S to X for a cash price of one hundred
dollars, and then to buy it back for a credit price of one hundred and
ten dollars payable next year.
The net effect is that the asset made a round trip from Y to X and back.
More importantly, the loan is synthesized since X gives Y one hundred
dollars now, and Y owes X one hundred and ten dollars payable in one
year.
This practice is actually quite common in Malaysia, where same item
sale-repurchase is not deemed forbidden, as long as the two sales are
not incorporated in a single contract.

However, the majority of schools of Islamic jurisprudence forbade same
item sale-repurchase without a third party intermediary.
Let the third party be $Z$, and assume that $Z$ is the party that owns S.
One credit sale and two spot sales now need to be conducted.
First, X buys S from $Z$ and pays one hundred dollars on the spot.
Second, X sells S to Y on credit, for a price of one hundred and ten
dollars, payable in one year.
Finally, Y sells S to $Z$ for the cash price of one hundred dollars.
Again, the net result is that the asset S has made a round trip: X paid
one hundred dollars now, Y received one hundred dollars now, and Y owes
X one hundred and ten dollars payable in a year.
Loan A has been synthesized again.
This transaction is commonly known as \emph{tawarruq}, which literally
means turning the asset S into silver, or monetizing it.
\end{quotation}
This quotation is concerned with financial products A, B and C where an
identity A = B + C may hold expressing that A is synthesized by
composing B and C.
From the point of view of formal methods in computing this text fragment
alone already leads to a range of questions.

It is not hard to find other sources about the synthesis of financial
products.
From Willem G. Wolters~\cite{Wol05a} we take the following quotation
(translated by us from the Dutch original):
\begin{quotation}
\noindent
Complex contracts emerged in which the direct imposition of interests
was avoided. A well-known example was the tripartite contract also
called \emph{contractus trinus}.
Instead of a loan from person A to person B for which interest was due
(which was forbidden) came three contracts each permitted by the Church,
which together had the same effect: a loan, a compensation and an
insurance.
In the first contract person A invests a certain sum of money (say
100 pounds) in a cooperation with person B (partnership contract) for
the time of a year.
In the second contract A sells to B a part of the profit of the
cooperation, for a fixed amount (say 15 pounds), to be paid by B.
In the third contract between A and B, A insures himself against the
loss of the principal sum by paying to B an amount, say 5 pounds.
The result is that A receives 10 pounds for a loan of 100 pounds for the
duration of one year.
This amounts to an interest of 10 percent per year.

In 1514 the German theologist Johannes Eck, working at the University of
Ingolstadt, declared the tripartite contract legitimate.
\end{quotation}

In the preceding examples, legal assessments come into play which
distinguish legitimate (\emph{halal}) from non-legitimate (\emph{haram})
contracts or progressions of actions involved.
These examples illustrate the simplest compositions of products only.
Because of their classical nature, we call them
\emph{fundamental synthesized product examples}.
Such examples have a long history and are in some sense the paradigmatic
examples of their kind.

The most obvious reason to seek a loan is the intention to purchase a
certain good while being unable to pay the full price in advance.
This decomposes in lending and buying the good with the loan.
The \emph{murabaha} contract has comparable result while avoiding the
payment of interest altogether.
Again we quote Wolters~\cite{Wol05a} (taking the liberty to use the name
G for the good referred to):
\begin{quotation}
\noindent
Person A intends to purchase a good G from B and asks the bank
for a loan.
The bank is willing to mediate in the transaction and asks A to promise
to buy G from the bank after the bank has bought G from B.
The bank is now certain that when it buys G from B it can resell G to A
immediately thereafter.
The bank agrees that A pays for the delivery after a certain time.
For the mediation, A pays the bank a compensation.
The bank subsequently buys G from B and immediately thereafter resells G
to A at a higher price (`cost-plus') that must be paid after a year.
\end{quotation}

For more detailed information on financial products of this kind we
refer to~\cite{ElG06a}.

\subsection{Reduced Product Set Finance}
\label{subsect-rpsf}

We have done extensive reading on the arguments for and against Islamic
finance.
We draw from that reading the conclusion that the investigation that we
bring up in this paper, and which we hope to carry out in the near
future, will have no bearing on those arguments.
The reason for this is that the point of departure of the investigation
will be the perception of Islamic finance as conventional finance
restricted by the prohibition of interest.

The very possibility (and fact according to some) that a significant
alternative to conventional finance can be found by means of a
restriction of the permissible financial methods is quite intriguing.
Moreover, obtaining an advantage from imposing design space restrictions
is a phenomenon known from computing.
In computer architecture, the limitation of instruction sets has been a
significant help for developing faster machines using RISC (Reduced
Instruction Set Computing) architectures.
Fast programming, as opposed to fast execution of programs, is often
done by means of scripting languages which lack the expressive power of
full-blown program notations.
Replacing predicate logic by propositional calculus has made many
formalizations decidable and for that reason implementable and the
resulting computational complexity has been proved to be manageable in
practice on many occasions.
New banking regulations in conventional finance resulting from the
financial crisis of 2008/2009 have similar characteristics.
By making the financial system less expressive, it may become more
stable and on the long run more effective.
Indeed, it seems to be intrinsic to conventional finance that seemingly
artificial restrictions are a necessity for its proper functioning.
The development of these restrictions is propelled by the drastic
innovations of the financial industry rather than by ethical constraints
of a principled kind.

We find it appealing to systematically study forms of finance where only
financial products that are synthesized from a few basic financial
products according to certain principles are considered.
We coin the term ``reduced product set finance'' for this kind of
finance.
We consider the form of finance that the avoidance of interest gives
rise to in Islamic finance the classical example of reduced product set
finance, but recognize that Islamic finance aims at objectives not
captured by reduced product set finance.

Restrictions other than the avoidance of interest have been imposed in
the past.
Mostly, the restrictions concerned the use of money.
Among them are the restriction that only state controlled agencies are
allowed to mint coins and the restriction that the issuer of coins keeps
the ownership of coins --- implying that holders of coins are not
entitled to extract valuable materials from coins.
The fact that interest rates are non-negative is a current restriction
which might be removed from conventional finance.
Redesigning the system without coins and banknotes using e-money only
will amount to yet another restriction which is bound to be studied in
depth in the near future.

Whatever the practical value of reduced product set finance, it
qualifies as an intellectual challenge at least to explore its
potential.
Our idea is to study the most prominent reduced product set finance
obtained by the prohibition of interest in a formal way.
Clarification of concepts and the analysis of proper distinctions is
considered prior to the assessment of economic, political and social
effects of putting a particular form of reduced product set finance in
place in a real economy.

\section{Impartial Investigation}
\label{sect-impartiality}

The question whether Islamic finance or aspects thereof can be
investigated from a non-Islamic perspective at all is a complex question
for which no brief and conclusive answer can be provided.
However, when writing from a non-Islamic background, simply by doing so,
we cannot deny to have a positive answer to that question in mind.
In view of the wide range of opinions that are being voiced about the
role of Islam in modern democratic countries, we believe that
investigating Islamic ideas from a non-Islamic perspective requires
taking up a position with regard to those opinions.
Below, we explain our viewpoints on this matter.
They support a context in which an impartial and systematic
investigation of aspects of Islamic finance can be pursued.
For none of our viewpoints we claim originality.
They derive in part from~\cite{Sal06a} and they can all be found in a
diversity of sources.

\subsection{Separation of Church And State}
\label{subsect-separation}

The objection that Islam fails to separate church and state and for that
reason lacks modernity is flawed.
The institutional separation of church and state and the ideological
separation of religion and politics are both processes which are still
under development in the West.
Western democracies may have been premature in declaring these issues
solved.
Confrontation with a substantial representation of Islam in EU countries
reveals weaknesses in the maturity of these separations.
Indeed this confrontation is likely to contribute to a further maturity.
There is room for the assumption that Western democracies have made use
of the fragmentation of Christianity to steer towards an equilibrium
which has not always been convincing from a religious (in particular
Christian) perspective.

\subsection{Integration of Islamic Minorities}
\label{subsect-integration}

The objection, nowadays often voiced in EU countries, that Islamic
minorities fail to integrate, thereby causing problems to their country
of adoption, cannot be held against Islam in any way.
Migration streams often lead to integration issues and this is by no
means specific for migration streams  from countries with Islamic
majorities.
There is no evidence that such minorities would be less productive in
terms of integration than minorities of other origin.

\subsection{Shariah and National Constitutions}
\label{subsect-shariah}

The objection that Islam promotes a vision that Shariah takes preference
over national constitutions, thereby being ``undemocratic'', cannot be
taken seriously, at least not as an objection specifically raised
against Islamic thought.
Christian and Judaic ideologies would prescribe similar viewpoints if
significant frictions with local constitutions arise, only stated in
different words.
If that form of independence would at some critical stage fail to
emerge, later generations of Christian communities might not praise
their ancestors for their obedience to time bound constitutions.

The confidence that Christians have in democratic processes is high but
limited in principle.
Their acceptance of these processes as leading is connected to an
implicit assumption that ``Christian values are dominant''.
The whole picture is turned upside down if one pretends that Christians
have all seen the light of democracy and now believe that always two
thirds of a population have ultimate authority on ethical matters.
This infallibility of the democratic process is of course equally
problematic as some similar principles that have been preached from
Rome.

Indeed a strong foothold of Islam in western democracies will force
Christians in those communities as much into carrying out an in depth
analysis of their appreciation of democracy as their Islamic
co-citizens.
If their appreciation of democracy is linked to the assumption that
Islam has a relatively weak position, then the maturity issues mentioned
above surface at once.
Otherwise, they have either embraced democratic mechanisms as
incorporating ultimate ethical values or they have recognized Islam as
a ``friendly force'' when put to the test in extreme cases.
Either way, these steps are quite non-trivial and Islam cannot be blamed
for the existence of the conceptual problems which are simply endogenous
to Western political thought.

\subsection{Import of Islamic Thought}
\label{subsect-import}

The import into Western Europe of Arabic and Islamic thought has been
massive, but it took place long ago and non-trivial historical awareness
is required to appreciate that fact.
Promoting that awareness is certainly not a major objective of the
current Christian-Judaic tradition.
Even the conception that Western Europe mainly stands in a
Christian-Judaic tradition may be contested.
Several rules that underly Islamic finance have been adopted in Western
finance as well.
Moreover, there is no reason to assume that an Islam tradition somehow
Europeanized may not flourish and have world wide impact.
There is no obvious ground for a specific and lasting commitment of
Western European countries to the oldest two of the three major
Abrahamic religions.
The mere argument that recent immigrations force several EU states into
the accommodation of Islam is besides the point.
Islam has never been very far away, and needless to say many of the
immigrants have been invited to come.

\subsection{Islamization of Modernity}
\label{subsect-modernity}

The main debate between the Christian-Judaic West and Islam seems to
concentrate on the interpretation and the advancement of modernity.
If Islamic influence implies an inflexible commitment to an explanation
of all social realities in terms of writings dating back some 1000 years
and more this very lack of commitment to open-minded innovation may
constitute  a major cause of frictions.
The well-known dichotomy between ``Islamization of modernity''%
\footnote
{From this ``Islamization of Science'' and ``Islamization of Finance'',
 subsequently resulting in Islamic finance, have been derived in the
 first half of the 20th century.
}
and ``modernization of Islam'' points to this issue.
If the more extreme viewpoint of ``Islamization of modernity'' is taken
as a point of departure, which may be compatible with an EU based
Islamic tradition with wold-wide ambitions, then the critical debate
will be about the form and objective of modernity as an ongoing and
progressive process.

\subsection{Competition between Christianity and Islam}
\label{subsect-competition}

If one denies the relevance of religion, it is obvious that
accommodating more rather than fewer religions is the way ahead while
defending some religions against the progress of other religions is hard
to explain.
If one accepts the potential relevance of religions, allowing a fair
competition between them is the way to go, while supporting a cartel is
not.%
\footnote
{Remarkably, the objection against cartels can be found in classical
 Islamic writings.
}
If one is convinced of the ideological superiority of Christianity or
Judaism, then there cannot be an objection against the investigation of
other religions and the political and economic impact of their tenets.
So its seems that our work is objectionable only from a perspective that
Christianity or Judaism needs protection in some Western democracies
because it lacks any convincing superiority and that in addition
investigating Islamic finance creates a problematic risk of infection.
By doing the work presented in this paper, we implicitly state that we
do not support the latter viewpoint.
That may be held against us and the consequences of such objections are
ours.

\section{General Questions}
\label{sect-questions}

Before the investigation proposed in this paper is conducted, answers to
certain general questions should be given.
Among these questions are the question whether the investigation can be
conducted without assumptions about the validity of ethical principles,
the question whether the investigation is timely, the question whether
the object of the investigation has an application perspective, and the
question whether the investigation is relevant to computer science.
In this section, we make an effort to answer the above-mentioned
questions.
We also go into the relevance of computer science to the investigation.

\subsection{Independence of Ethical Principles}
\label{subsect-ethical-principles}

The investigation will not be based on any assumption about the validity
of ethical principles such as the prohibition of interest payment and
charging.
Independently of any such assumption, it can be observed that a very
substantial community considers such forms of prohibition important.
We will focus primarily on the phenomenon of interest and ways of
avoiding its usage.
Because no ethical objections have been published against such forms of
prohibition, there seems to be no conceivable objection against taking
it very seriously as a potential design decision of a modern financial
system and investigating its role and meaning in technical terms.
Our work can be understood in this line.

To our knowledge, a system of ethical banking based on the prohibition
of interest as well as some other well-known restrictions is a subsystem
of conventional banking.
Investigation of the expressive power of a limited subsystem has always
been considered important in computer science and the line of thought
that we propose to follow is similar in spirit.

\subsection{Timeliness}
\label{subsect-timeliness}

One may consider the issues to be investigated futile because the
prohibition of interest constitutes a step backward from a conventional
point of view.
Against that viewpoint one may put forward that, given the extreme
importance of interest rates in conventional finance, it is an essential
thought experiment to understand in detail what happens if that
mechanism is ruled out.
The fact that this thought experiment has an extremely longstanding
history does not imply that all of its consequences have been completely
investigated.
We suggest that formalization techniques that have been developed in
computer science may still provide interesting and new insights in this
very classical area.

So rather than asking what is lost by the prohibition of interest, one
may ask what is gained by permitting interest from a situation where it
is not being used.
In other words, a reduced product set finance may lead to an effective
system for which the addition of some feature may be less advantageous
than expected by those who take the feature for granted.
Besides prohibition of interest charging and payment, the prohibition of
excessive and unnecessary downside risk (\emph{gharar}), see
e.g.~\cite{ElG01a}, is an equally interesting limitation of financial
products, though far harder to study because of its conceptual
difficulty.
Indeed, one cannot require the absence of uncertainty in all
circumstances and removing \emph{gharar} seems to require a mixture of
qualitative and quantitative reasoning.

For those who do not see the point of interest prohibition at all, we
recommend~\cite{Lew99a}.
That paper provides a harsh survey of the practice of credit provision
in the US, which suggests that at least in some cases prohibition of
interest may be a plausible definite step forward.
The problems inherent in a liberal market place based on credit and
interest are very convincingly put forward.

Mohammad N. Siddiqi states in~\cite{Sid02a} that more fundamental
research on Islamic finance is welcome.
The investigation proposed in this paper may add to such work.
However, whether our points of departure will qualify as points of
departure for such fundamental research in the perception of Siddiqi is
another matter.
Rania Kamla~\cite{Kam09a} suggests research on Islamic accounting to
move far beyond interest prohibition.
His paper convincingly criticizes the position that interest prohibition
is the most central distinction between Islamic finance and conventional
finance.
Indeed we do not suggest that an investigation of the foundations of
interest avoidance and interest prohibition are necessarily high
priorities in a modern research agenda on Islamic finance because many
other issues may be of equal or higher importance.

Statistical work on financial techniques applied in Islamic banking as
well as the geographical distribution of Islamic banking can be found
in~\cite{BDW10a,CL09a}.
The results of such work must also be taken into account when making an
assessment of the timeliness of fundamental research in interest
prohibition for Islamic finance at large.
In the light of the above-mentioned papers, we conclude that the
priority of this research is moderate at best.

\subsection{Application Perspective}
\label{subsect-perspective}

As outlined in Section~\ref{subsect-timeliness}, there seems ample
justification for detailed investigations of interest free forms of
reduced product set finance.
However, one may still hold that interest free finance is an incoherent
concept per se and that it has no application perspective.
Many objections have been put forward and many papers contain thorough
responses to a selection of such objections.
We mention~\cite{Cha85a} as a systematic discussion in this respect.
However, there seems to be a number of questions about the rationale of
interest free forms of reduced product set finance which are hardly
taken into account by its proponents, to such an extent that this
constitutes a risk for the application perspective of interest free
forms of reduced product set finance.
Here we list some of these questions:%
\footnote
{In \cite{Iqb10a} some of these arguments are taken care of and an
 entirely different path of argument is taken: it is argued that
 prohibition of interest, or rather equity based financing, is a
 rational mechanism which minimizes risk in some mathematically precise
 way.
 While providing grounds for interest prohibition, that argument seems
 not to prove the necessity of interest prohibition.
 So the question remains why the optimization result with regard to
 risk would give rise to a strict normative judgement.
}
\begin{itemize}
\item
Assume that $X$ may borrow an amount $p$ from $Y$ during time interval
$I$ without paying interest.
How should $Y$ be compensated for the risk that $X$ defaults and fails
to return the amount $p$ at the end of $I$?
One might say that the presence of this risk constitutes \emph{gharar}
and for that reason it need not be taken into account.
That is, $Y$ should not engage in the transaction if there are worries
about $X$'s ability to return the principal sum.
That by itself is significant, as it would point out that interest
free reduced product set finance needs to incorporate additional
restrictions.
But the argument is weakened by the fact that the risk of default may be
quite low which is an indication against \emph{gharar}.
\item
The significance of prohibition of interest is far more easily
understood in the context of a firm and stable metallic (or bi-metallic)
standard than in a setting where money is sensitive to inflation and
deflation.
As it seems to be the case that classical authors all had some metallic
standard in mind, or even made no distinction between money and precious
metals, one may question the very proposition that their viewpoints were
meant to be applied to all future forms of money, including modern fiat
money.
In general, there is hardly enough analysis of the concept of money when
prohibition of interest is proclaimed.
\item
Proponents of an interest free reduced product set finance often point
out the disadvantages of interest payment as a means of conducting
financial business.
It is hard to understand why a mechanism which may sometimes, but not
always, lead to adverse consequences must be entirely forbidden.
If there are disadvantages correlated with the usage of interest, one
needs policies to remedy these disadvantages.
One cannot but conclude, so it seems, that today's proponents of
interest free banking and finance base their judgement exclusively on
the interpretation of classical legal scholarly writing within a variety
of religious traditions.
This form of analysis induces an uncompromising judgement on absolute
moral grounds which is not amenable to exceptions.
\item
If $X$ provides a loan $p$ to $Y$, one may hold that $X$ provides a
service to $Y$.
But equally well $Y$ may be providing the service to $X$, either because
$Y$ can store $p$ for $X$ or because collecting interest on the loan is
the best or even only method available to $X$ to get any rewards from
its possession of the money.
More attention should be paid to the variety of motives for $X$ and $Y$
for a seemingly similar transaction when contemplating judgements of
legality.
\item
The strongest argument against a full prohibition of interest is that
the profit-loss contracts which are advocated instead are likely to have
far higher transaction costs than interest compensated loans.
Indeed if $X$, rather than borrowing $p$ to $Y$, provides $p$ to $Y$ for
combining forces in a joint enterprise that is mainly managed by $Y$,
whereas $X$ is entitled to some fraction of its revenues, $X$ inevitably
is dependent on being correctly informed about $Y$'s results.
Even worse, $X$ and $Y$ seem to have conflicting interests (in the other
sense of the word).
The simplification introduced by a loan with interest, with respect to a
profit-loss sharing participation in another party's enterprise, is that
a complete correspondence of objectives between both parties is obtained
if $X$ is not fraudulent (otherwise both options are unsatisfactory
anyhow).
It follows that, in the case of a loan with interest, $X$ can manage his
part of the contract on the basis of far more abstract information about
$Y$.
The information in question may be obtained in a less intrusive fashion,
which is profitable for maintaining proper relations between $X$ and
$Y$.
Of course, this argument would disappear if the prohibition of interest
was judged on a case by case basis.
However, such flexibility is absent in recent writings on interest
prohibition.
\end{itemize}
Many answers on these questions and combinations of them can be
imagined.
It cannot be claimed that these questions need to be resolved
unambiguously for a research project in interest free reduced product
set finance to be justified.
It can hardly be denied, however, that any application of the results of
research in this area will critically depend on the collection of
answers that are provided to these objections to the prohibition of
interest.

\subsection{Relevance to Computer Science}
\label{subsect-relevance-to-cs}

Besides suggesting that financial product synthesis merits substantial
further investigation, we also claim that doing so may be performed by
means of methods and techniques which were originally developed in
computer science.
In principle, a proof of this claim can only be given by actually
carrying out the work with the help of such methods and techniques.
However, it is meaningful to speculate about the possibility that
results may be fed back to the computer science setting.
Below, we give some grounds on which we expect this to be a realistic
possibility.

\subsubsection*{Behaviours with promises and obligations}
Probably the simplest connection between computing and the world of
financial product synthesis is found if one compares a contract with a
control code or an instruction sequence as found in computing.
The presence of a plurality of contracts between different agents
suggests the presence of some form of multi-threading.
Moreover, promises and obligations play an important role.
Specifying the behaviour of computer-based systems in terms of promises
and obligations is relatively new, however, and progress may be obtained
by making use of the results of a systematic analysis of financial
products.

\subsubsection*{Classification of artifacts}
In computing, the classification of large collections of software
artifacts in a small number of categories takes place quite often while
theoretical work supporting such classifications is much less developed.
A plan for reverse application into computer science is that solutions
to the classification problem concerning financial product legality may
shed light on classification problems which can be found in computing.
Here we mention some of these latter classification issues:
\begin{itemize}
\item
Control code (see~\cite{BM07b,Ber10b}) can be classified as either
programmed or non-programmed (acquired by dark programming in the
terminology of~\cite{Jan08a}).
More conventionally control code can be classified as either executable
or non-executable.
Both classifications are far from obvious, in spite of their intuitive
appeal.
\item
Control code may also be marked as malicious.
Again the intuition is clear but a formalization of this property is
hard to provide.
\item
Computer programs are often understood as being either faulty or correct.
This distinction is somehow comparable to the distinction between illegal
and legal in the the case of financial products.
Moreover, the connection between program correctness and program testing
is difficult to assess because software testing has not been defined
clearly (see~\cite{Mid10b}).
\item
The role of control code in a system may be hard to assess.
For example, it is difficult to classify control code as either system
software or application software.
It is frequently mentioned that operating systems belong to system
software, but it has been pointed out in~\cite{Mid10a} that the very
classical notion of an operating system fails to have been provided with
a clear definition up to now.
\item
In a computer-based system some of its concurrent processes may be
threads (see~\cite{BM04c,PZ06a}).
The intuition is that all threads are processes but not conversely.
Which processes are threads is a matter of definition, but there is no
consensus about the definitions in question.
The abundant use of the terminology of multi-threading seems to stand in
the way of a proper understanding of this matter.
\end{itemize}

\subsubsection*{Computer support for financial products}
Computerized financial systems are among the most important applications
of computers.
It is reasonable to assume that an improved understanding of financial
products can also be helpful for an improved understanding of computer
support for such products (see e.g.\ the literature on Islamic credit
cards~\cite{Bak06a,NA09a}).

\subsubsection*{Certification of compliance with ethical values}
If one assumes that legal issues concerning financial products derive
from the intention to design and to improve systems for ethical banking
and finance, this intention by itself may be meaningful for the purpose
of the design of computer-based systems.
Increasingly, ethical issues appear to be central to the design of
computer-based systems.
It may be quite important to be able to certify that a computer-based
system incorporates some ethical values that have been specified in
advance.
Doing so is far from trivial and making use of examples from other areas
may be of help.

\subsection{Relevance of Computer Science}
\label{subsect-relevance-of-cs}

We expect to use multi-threading as an operational paradigm for putting
financial products into effect, instruction sequences as considered in
program algebra as a means for defining threads for parts of financial
products whenever possible, and quantities from a meadow as a means for
measuring money.
Below, we give grounds on which we expect this.

\subsubsection*{Multi-Threading}
Financial products, when put into effect, nearly always involve the
activity of different agents.
For instance, a bank is considered an agent.
The different agents operate concurrently.
Each agent on its own acts in a sequential fashion, usually executing a
predetermined plan.
Therefore, it is plausible to think of the behaviour of an individual
agent as a thread.
The concurrent composition of several threads needs to guarantee that
each agent is able to perform the planned actions in a regular fashion,
without some agents being starved due to repetitive and uninterrupted
activity of other agents.
Precisely this kind of parallel cooperation is captured by
multi-threading for which we have developed very elementary
formalizations in~\cite{BM04c}.

\subsubsection*{Instruction Sequences}
The use of instruction sequences as considered in program algebra
(see~\cite{BL02a,PZ06a}) as a means for defining threads for parts of
financial products is based on the observation that in computing an
instruction sequence is a very plausible means for defining a thread.
However, we do not claim that all contracts can be decomposed into a
number of threads of which each is defined by means of an instruction
sequence that captures the essence of a part of the contract.
Indeed contracts may be too sophisticated or too involved to admit such
a decomposition.

\subsubsection*{Meadows}
Meadows (see~\cite{BT07a,BR08a}) are appropriate mathematical structures
for quantities if quantities are measured with finite accuracy, as in
the case of money.
The prime example of meadows is the field of rational numbers with the
division operation made total by imposing that division by zero is zero.
Meadows are mathematical structures to which techniques developed for
abstract data types in computer science can be applied.

\section{More on Synthesized Financial Products and Legality}
\label{sect-more-synthesis}

We believe that, in order to achieve reliable results, the formalization
of financial products and principles of financial product synthesis must
form part of the investigation of the main issues concerning reduced
product set finance.
That formalization requires a clear understanding of synthesized
financial products and the prevailing positions concerning their
legality.
In this section, we pose questions about the fundamental synthesized
product examples from Section~\ref{subsect-examples-synthesis} and the
prevailing legal positions that need answers in order to gain a clear
understanding.
Moreover, we attempt to give a provisional answer to the question which
actions or events should be distinguished when formalizing the
fundamental synthesized product examples and list a number of actions
which may not qualify as ethically correct from the perspective of
certain legal positions.

Henceforth, in the context of financial products, the term ``state'' is
used to refer to a snapshot of the conditions in which a financial
product is when it is put into effect and the term ``progression'' is
used to refer to a succession of actions that results from putting a
financial product into effect.

\subsection{Questions about Synthesized Financial Products}
\label{subsect-questions-products}

In spite of their respectable age of several centuries, a number of
questions can still be posed about the fundamental synthesized product
examples from Section~\ref{subsect-examples-synthesis}.
We list a number of questions that occurred to us:
\begin{itemize}
\item
Which catalogue of legitimate basic financial products, services and
contracts can be assumed for further synthesis?
We mention that a list of \emph{halal} products, with informal
specifications of each of them, can be found in~\cite{GW09a}.
\item
Which categories of objects need to be distinguished, and what
definitions are to be used for them?
Possible categories include the categories of states, agents, accounts,
goods, values, contracts, binding promises~\cite{AlM02a}, non-binding
promises~\cite{Bur07a}, progressions, threads~\cite{BM04c}, and
processes~\cite{BBR10a}.
\item
Which actions or events should be distinguished when formalizing the
fundamental synthesized product examples?
\item
What is the semantic complexity of product legality?
That is, is legality of a product based on an assessment of one of the
following:
\begin{itemize}
\item
one or more contracts;
\item
a single progression;
\item
a tree of progressions that forms the behaviour of a thread;
\item
a more general process behaviour;
\item
a particular combination of entities of the above-mentioned kinds?
\end{itemize}
\item
Which principles of financial product synthesis are used in the cases of
the fundamental synthesized product examples?
\item
Can it be taken for granted that synthesis of products from legitimate
products leads to legitimate synthesized products?
If not, why can it be assumed in the fundamental synthesized product
examples?
In other words, what are valid principles of synthesis?
We note that special attention must be paid to the fluctuations in value
of the asset or business involved between the different actions or
events that make up the fundamental synthesized products.
\item
Is it possible to design a large space of potential products closed
under certain principles of financial product synthesis that contains
both legal and illegal products (according to some definition of
legality), including the fundamental synthesized product examples?
\item
Can judgements about legality, based on some set of general assumptions,
be automated in the sense that decisions about legality can be made by
means of an unambiguous logic?
And if so, in which formalism can the conditions that determine the
legality of a contract (according to a specific legal position) best be
expressed?
Some candidates are progression ring notation~\cite{BP09b}, (timed)
tuplix calculus~\cite{BM09a}, thread algebra~\cite{BM04c}, process
algebra~\cite{BBR10a}, temporal logic~\cite{KM08a}, predicate
logic~\cite{HR04a}, and propositional logic~\cite{HR04a}.
\item
Are there cases where legality of a product is not determined by its own
structural properties but rather by the use that is made of it in a
specific context of agents.
In other words: must the relevance of the use of the product for one or
more of its participating agents be taken into account when deciding
about its legality?
\end{itemize}

\subsection{Questions about Legal Positions}
\label{subsect-questions-positions}

Different legal positions concerning product legality occur in practice.
We list a number of questions about those legal positions that occurred
to us:
\begin{itemize}
\item
Can a comprehensive and comparative survey of the legal positions
about financial products be given?
\item
How can a legal position be specified?
\item
For each legal position (given the presence of a finite number of
agents, say~$k$):
\begin{itemize}
\item
What counts as parameter of financial products (in addition to the saver
or borrower and the principal sum deposited or borrowed)?
\item
How many legitimate financial products exist (finitely or infinitely
many, and if only finitely many: how many as a function of $k$)?
\item
Which symmetries can be found, and must roles of agents be distinguished
(e.g.\ banks, companies, and private individuals)?
\item
Is it meaningful to allow auxiliary agents, and how does that impact the
counting of legitimate products?
\end{itemize}
\item
Is there a notion of completeness for a legal position?
And if so, is a finite presentation of a complete legal position
necessarily equipped with rules that have infinitely many instances?
\item
Are there well-definedness criteria for a legal position?
Examples of well-definedness criteria are that legality must depend on
behaviour only and that legality must depend on externally observable
behaviour only.
\end{itemize}

\subsection{Further Thoughts on Legal Positions}
\label{subsect-thoughts-positions}

It is an intriguing fact that according to some legal positions very
substantial differences exist in the legality of seemingly related
financial products or compositions thereof.
Taking a semantic view from computer science as a point of departure, it
must be admitted that initially no intuition about how to recognize
these differences is given.
The design space of potential financial products, atomic and composite,
needs to be enriched with design requirements and rules in order to
single out legitimate designs.
Finding out about those requirements and rules must be done by asking
questions to experts or, in the presence of an abundant literature, by
reading professional texts about these legality judgements.
The following thoughts may be important to this matter:
\begin{itemize}
\item
If no clear requirements and rules can be formulated, machine learning
can be employed to detect the patterns which are considered decisive for
legality.
\item
A qualitative question for each legal position is whether its judgements
are binary or there exists a larger range of judgements.
If binary judgements cannot be achieved, the question arises how to
decide about the particular cases where a binary judgment cannot be
achieved.
\item
If clear patterns can be found, specifications of legal positions can be
provided of which each comprises a sound and complete system of
inference for product legality.
\item
Only after the fundamental composite product examples have been
completely understood, one should turn attention to more complex
products involving mortgages, pension funds, insurances, shares, bonds,
options, futures, swaps and so on.
\item
An investigation of product synthesis can be helpful if certain aspects
considered relevant to legality are to be better understood.
Here, one may think of products which feature an excessive downside
risk, products which when put into effect provide asymmetric information
to those who are involved in a problematic fashion, products which when
put into effect are prone to force one or more of those who are involved
into undesired actions, and products which may invite usury to an
excessive extent.
\end{itemize}

\subsection{Actions Involved in the Use of Financial Products}
\label{subsect-actions}

The use of financial products and services involves actions (also called
events), such as speech acts (promises, justifications, claims,
acknowledgements, threats), transfers, sales.
The following are some actions that should be distinguished when
formalizing the fundamental synthesized product examples:
\begin{itemize}
\item
$X$ promises $Y$ to pay a sum of money $p$ to $Y$ at date $d$ for
reasons $r$;
\item
$X$ promises $Y$ to accept $Y$'s payment to him of a sum of money $p$ at
date $d$ for reasons $r$;
\item
$X$ makes a payment of a sum of money $p$ to $Y$ via channel $m$ at date
$d$ for reasons $r$;
\item
$X$ receives a payment of a sum of money $p$ from $Y$ via channel $m$ at
date $d$ for reasons $r$;
\item
$X$ acknowledges to all agents involved in a financial product that he
has received a sum of money $p$ from $Y$ at date $d$ for reasons $r$;
\item
$X$ promises $Y$ to buy a product or service $P$ from $Y$ as soon as a
certain condition $\phi$ is met;
\item
$X$ buys $P$ at price $p$ from $Y$ at date $d$ and promises to pay $p$
at a later date~$e$;
\item
$X$ asserts that he expects a sum of money $p$ from $Y$ before date $d$
for reasons~$r$;
\item
$X$ justifies with reasons $r$ why he is entitled to receive a sum of
money $p$ from $Y$ at date $d$;
\item
$X$ promises $Y$ to pay a sum of money $p$ to $Y$ at date $d$ if he
receives a sum of money $q$ for an insurance policy to that extent
before date $d$;
\item
$X$ promises $Y$ to manage a sum of money in the range $p$ to $q$ for
$Y$ from date $d$ to date $e$, under the condition that $Y$ transfers
the sum of money in question before date $d$;
\item
$X$ exchanges a sum of money $p$ with $Y$ at date $d$, where $p$ is
available in different coins and/or banknotes.
\end{itemize}
We note that a promise produces a signed contract of a certain form.
There are many degrees of freedom in the design of contracts.
For instance, the contract is signed by $X$ and $Y$ or only by $X$, a
partially or wholly unsigned version of the contract is available before
it is signed or not, and descriptions of reasons are themselves included
the contract or only references to these descriptions are included.
Because of this, the possible promises are numerous.

\subsection{Ethically Incorrect Actions}
\label{subsect-actions-unethical}

From the perspective of certain legal positions, actions do not qualify
as ethically correct if they involve speculation, information asymmetry,
coercion or interest.
The following are examples of such actions:
\begin{itemize}
\item
$X$ promises $Y$ to pay a sum of money $p$ at date $d + 2$ if it is
raining in place $l$ at date $d$;
\item
$X$ sells a used automobile $A$ to $Y$ at date $d$ for a sum of money
$p$, transferred by $Y$ in cash, while $X$ knows about technical
problems with $A$ not revealed to $Y$ before the transaction is made;
\item
$X$ asserts that he will severely damage $Y$'s possessions unless $Y$
pays a sum of money $p$ to him as a payment for a marginal or even
unwanted service $S$;
\item
$X$ pays a sum of money $p - c$ to $Y$ at date $d$ with the
justification that:
\begin{quote}
$Y$ has promised $X$ to pay a sum of money $p + i$ at a later date $e$
if he has received a sum of money $p - c$ from $X$ at date $d$ or
before, where $c$ represents the costs for $X$ of providing the loan and
$i$ represents the opportunity costs for $X$ of lending the sum of money
$p$ during a time period of length $e - d$ to $Y$;
\end{quote}
\item
$X$ receives a sum of money $p - c$ from $Y$ at date $d$ with the
justification that:
\begin{quote}
$X$ has promised $Y$ to pay a sum of money $p + i$ at a later date $e$
if he has received a sum of money $p - c$ from $Y$ at date $d$ or
before, where $c$ represents the costs for $Y$ of providing the loan and
$i$ represents the opportunity costs for $Y$ of lending the sum of money
$p$ during a time period of length $e - d$ to $X$.
\end{quote}
\end{itemize}

\section{Anatomy of a Noncompliant Financial Product}
\label{sect-loan-with-interest}

The objective to develop a compositional theory of financial products is
a distant target.
Indeed the design of non-compositional descriptions will precede
successful compositional methods.
Thus, a first priority for the development of a compositional theory is
to obtain semantic models for very simple products like a spot sale,
a credit sale, a loan without interest, and a loan with interest.
Below we will focus on the last product, as the most prominent example
of a financial product.

From the perspective of a given legal position, compliance of a
financial product can be assessed.
Some very well-known financial products are considered non-compliant
from the perspective of some very well-known legal positions, in
particular the classical legal positions which came about in a stable
form from the internal proceedings of the mainstream Abrahamic religions
mainly prior to the reformation of Christianity.

However clear this state of affairs may be at first sight, from the
perspective of formalization, the question must be posed how much
information regarding some financial product must be provided to
generate a judgement of non-compliance in a reliable way.
This question, which can be posed about all potentially non-compliant
financial products, including the most elementary ones, is about
definitions.
The issue is how financial products such as a loan with interest must be
formally defined.
In the remainder of this section, we look at a loan with interest with
the purpose to get a preliminary picture of how it must be defined.
We are looking for a definition in the strict sense of a definition that
provides a semantic model.%
\footnote
{The meta-theory of definitions as given in~\cite{Ber10a} will implicitly
 be used.
}

\subsection{A First Definition of a Loan with Interest}
\label{subsect-first-def-lwi}

One might say that, if the legal position is ``Islamic finance'',
without further specification of some specific direction of thought, and
the financial product is defined as ``taking a loan that carries
interest, including the action of interest payment'', we have an example
of non-compliance.
But this is not the insight that we are after.
The problem lies in the far to great dependence on the meaning of the
terms loan and interest.

\subsection{Splitting a Loan with Interest into Two Transactions}
\label{subsect-splitting-lwi}

More detail is given in the following description.
A loan with interest comprises the following elements:
\begin{itemize}
\item
A borrower $X$ and a lender $Y$, both assumed to be persons for
simplicity. $Y$ offers a service to $X$ by enabling $X$ to use a
quantity of $Y$'s money, called a loan, during a certain time period.
\item
A principal sum $p$, which is a quantity of money.
\item
Dates $d$ and $e$, with $d$ before $e$, such that on date $d$ a sum of
money $p - c$ is transferred from $Y$ to $X$ and on date $e$ a sum of
money $p + i + c'$ is transferred from $X$ to $Y$.
Here,
$i$ represents the opportunity costs for $Y$ of lending the sum of money
$p$ during the time period $e - d$,
$c$ represents the costs for $Y$ of providing the loan, and
$c'$ represents additional costs for $Y$.
\end{itemize}
A definite weakness of this description of a loan with interest lies in
the unclarity about the relation between the two transfers.
Is the relation a causation, a relation of justification, a relation of
enabling or a relation via a unique third object to which both transfers
refer, such as a signed contract whose preparation precedes the two
transfers?

\subsection{A Purely Contractual Definition of a Loan with Interest}
\label{subsect-contractual-lwi}

In a purely contractual view, the borrower and the lender both sign a
single contract in which they promise to make the two transfers with
appropriate references to the contract.
Implicitly that contract contains the concept of interest payment,
though no transfers are involved in a contractual definition.

The difficulty with a viewpoint where the legality of contracts is
assessed instead of the legality of progressions is the fact that
progressions do not follow unambiguously from contracts.
If several related legal contracts are partially honoured, it cannot
be excluded that the resulting progression coincides with the most
plausible progression that may arise when honouring an illegal contract.

\subsection{Aspects of a Semantic Model of a Loan with Interest}
\label{subsect-sem-aspects-lwi}

A semantic model of a loan with interest has to cover the following
aspects:
\begin{itemize}
\item
at ground level, the state of a loan with interest is made up of agents,
accounts, ownerships of accounts by agents, sums of money on accounts,
and perhaps goods and ownerships of goods by agents;
\item
on top of the ground level, there is a contract level at which there is
a collection of contracts at certain stages in their life-cycle;
\item
on top of the contract level, there is a plan level at which there is a
plan of future behaviour for agents based on contracts;
\item
on top of the plan level, there is a history level at which there is a
selection of historic information about preceding states at the other
three levels;
\item
transitions from one state of a loan with interest to another take place
as a result of actions performed by agents.
\end{itemize}
There are links between elements from the different levels that need to
be maintained, e.g.\ the links between planned actions and the contracts
that give occasion to them.
The following marginal notes with regard to the plan level are in order.
If several agents are involved in future behaviour, their joint future
behaviour is a matter of multi-threading.
The plan of future behaviour may change from state to state.

The above-mentioned aspects are by no means specific for a loan with
interest.
However, we claim that a semantic model already has to cover them in the
case where the financial product is a loan with interest.

The semantic model of a loan with interest explicitly or implicitly
includes the relevant sequence or sequences of states (sequences if
putting the loan with interest into effect can be done in several ways).

Given the above-mentioned aspects, it is convenient to look upon a loan
with interest as a transition system or an abstraction thereof.
This allows for a process, a thread or a progression, each constituting
simplified and abstract forms of transition systems.
The states of the transition system consist of four components, one for
each of the above-mentioned levels.
Appropriate abstractions can be given in terms of equivalence relations
on transition systems.
Because the states are not fully abstracted from in the abstractions in
question, these equivalence relations are reminiscent of the one found
in process algebra with signals~\cite{BBR10a}, where signals play the
role of states.

If the transition system is cycle free, histories up to some state can
be reconstructed unambiguously, which makes the fourth level redundant.
However, with all transition systems but the smallest ones, it is
probably significant to make use of a history level and to do away with
the notion of a history as formed by a path from the root through the
transition system.

\subsection{A General Definition of Interest Is Elusive}
\label{subsect-def-interest}

In spite of the progress made above, the very concept of interest has
many facets and a single definition that covers all instances as a
special case seems to be absent.
This state of affairs is very well-known in the theory of computing.
For instance the subject of parallel computing leads to many different
points of view, notations and semantic models of parallel computation,
with no model being the ultimate most general one.
The difficulty of providing a general abstract conception of interest,
say paid by $X$ to $Y$, comes from the many degrees of freedom, each of
which requires a choice to be made for.
Among these degrees of freedom are:
\begin{itemize}
\item
$X$ provides a service to $Y$ or conversely;
\item
$X$ is stronger in terms of wealth and power then $Y$ or conversely;
\item
$X$ takes the initiative or $Y$ does so;
\item
$X$ is in need of money or $Y$ has an excess of money;
\item
$X$ pays interest to $Y$ at the same time as he repays the principal sum
to $Y$ or $X$ pays interest to $Y$ periodically;
\item
$X$ knows the risk that he defaults and fails to repay the principal sum
to $Y$ or $X$ does not know this risk;
\item
$Y$ knows the risk that $X$ defaults and fails to repay the principal
sum to him or $Y$ does not know this risk;
\item
$Y$ splits the risk that $X$ fails to repay the principal sum from the
risk that $X$ fails to pay interest or he does not split these risks;
\item
there exist different ways in which $Y$ can cover the risks;
\item
there exist different ways in which $Y$ can optimize the probability
that $X$ will be able to repay the principal sum to him.
\end{itemize}

It may be concluded that a definition of interest has only a chance with
a fully specified example of a loan with interest instead of working
with an abstract example.

\subsection{Borrower in Need of Credit}
\label{subsect-borrower}

We consider the case where $X$ borrows a sum of money $p$ from $Y$ and
$X$ needs the sum of money.
Although this seems to be an obvious case of loan with interest, there
seem to be significant problems:
\begin{itemize}
\item
$Y$ will not lend $p$ to $X$ unless $Y$ has found evidence that $X$ is
able and willing to borrow $p$ from $Y$.
\item
$X$ and $Y$ must agree on some contract which both have acquired from a
third party $Z$, a broker, who provides example contracts and mediates
in a deal.
\item
$X$ takes the initiative by looking into the portfolio of contracts
which $Z$ has on offer and selecting a contract.
Then $X$ signs a contract, say $C$, which he considers applicable, sends
it to $Z$, who approaches $Y$ and finds out if $Y$ is willing to sign
$C$ as well.
If $Y$ is not willing to sign $C$, this is communicated by $Z$ to $X$
and there is no deal on the basis of $C$, after which $X$ may either
look for a different contract from $Z$'s portfolio, a different lender
or a different broker, or give up on obtaining a loan altogether.
If $Y$ signs $C$ as well, a signed copy is returned to $Z$ who
(i)~sends a copy to $X$,
(ii)~informs $Y$ about time, method and amount of payment to $X$, and
(iii)~informs $X$ about the time of payment, so that $X$ can warn $Z$ if
no payment has been received by that time.
If $X$ has to warn $Z$, $Z$ will make a claim to $Y$ on behalf of $X$
and there is no deal.
If $X$ receives the money in time, $Z$ prepares to inform $X$ later
about time, method and amount of repayment.
\item
The contract $C$ will specify how $Y$ will obtain guarantees that $p$
will be repaid.
There are several possibilities:
\begin{itemize}
\item
$X$ may hand over to $Y$ some valuable good which has a market price
above $p$.
$X$ permits $Y$ to sell these valuable goods if the money is not paid
back on time.
$Y$ is not permitted to use the valuable goods for any other purpose.
\item
$X$ may possess some valuable goods which are handed over to $Y$ once
$X$ fails to return the principal sum on time.
In this case, the difficulty is that $Y$ must in addition see to it that
$X$ will not sell these goods earlier, leaving $Y$ without guarantees
after all.
\item
$X$ may have an income which can be inspected by $Y$ and which $Y$, as
the need arises, can claim a share from until debts have been repaid.
In this case, it is a complication that the income may be lost before
the end of the planned period of the loan.
\end{itemize}
Each of these possibilities is different and broker $Z$ must offer
different contracts for each of them.
It seems that the first possibility is the simplest.
However, there is the complication that $X$ needs to obtain guarantees
that $Y$ will return the valuable goods upon $X$'s repayment of the
loan.
The contractual side of such guarantees is complex.
If $X$ does not insist on such guarantees, it is far simpler for $X$ to
sell the valuable goods instead of taking a loan.
\end{itemize}

\section{The Example of a Savings Account with Interest}
\label{sect-savings-account-with-interest}

Even the simplest example of a borrower in need of a loan as described
in Section~\ref{subsect-borrower} is quite complex.
The complexities in question may even defeat credible attempts to
formalization.
For that reason, we will now consider a simpler example where the lender
has an excess of money.

\subsection{Outline of a Savings Account with Interest}
\label{subsect-outline-sawi}

We suppose that person $X$ has obtained a gift or inheritance consisting
of valuable goods which he can sell on a market.
Because $X$ has no need of the goods and storing them is a burdensome
task, he sells the goods for a sum of money $p$.
Now $X$ must store $p$ until he wants to make use of it.
Doing so himself creates a risk of theft or loss due to misfortune, and
for that reason $X$ intends to find a bank $Y$ who will take $p$ as a
loan and repay this sum after an agreed time.
Because $p$ is not too high, the ability of $Y$ to repay this principal
sum is guaranteed by the state and $X$ does not need to worry about that
problem.

$Y$ offers to repay $p - c + i$ instead of $p$, where $c$ are low
transaction costs independent of $p$ and $i$ is defined as a fraction of
$p$, with the argument that
(i)~it can apply $p$ to a portfolio of useful projects which on average
produce significant returns on investment,
(ii)~it considers it fair to allow $X$ some share in that profit,
(iii)~a competing bank, say $Y'$, offers to repay $p - c + j$ where
$j > i$, but $Y$ claims that this promise may induce a risk that $Y'$
fails and that the subsequent process $X$ needs to follow to make use of
state guarantees is quite unattractive,
(iv)~the fractional-reserve banking system allows it to produce loans to
other clients in need of money on the basis of the fact that it has
obtained the deposit $p$, and in this way depositing $p$ may be
considered a contribution of $X$ to the local community which merits a
reward.

The borrower (bank) thus offers a compensation for money put on deposit.
In economic terms, $Y$ is the stronger agent and in addition to this he
has the backing of the state to guarantee his obligations towards $X$.
In this case, it is quite hard to see why it should be forbidden that
$Y$ repays $p - c + i$.
In this scenario, it is quite plausible that $X$ has a surplus of money
available over the sum $p$ he intends to deposit.
We will make use of that fact when considering the options for a
synthesis of this loan from permissible products.

\subsection{Details for a Savings Account with Interest}
\label{subsect-details-sawi}

The case of a lender $X$ who has a surplus of money for which he asks a
borrower $Y$ to keep it in storage requires many details for its
formalization:
(i)~a broker $Z$ should produce model contracts from which $X$ can make
a selection,
(ii)~$Z$ mediates the communication between $X$ and $Y$ leading to a
contract signed by both $X$ and $Y$ with copies available to $X$, $Y$
and $Z$, and
(iii)~$Z$ can apply for law enforcement if $Y$ fails to repay the
principal sum plus an agreed additional sum.

The model contract which $X$ obtains from $Z$ after having made a
selection should be a model in which the principal sum $p$ can vary over
some significant interval.
By providing a formula $p - c + q \cdot p$, where $c$ and $q$ are
constants, to determine what has to be paid back by $Y$ to $X$ after a
given time period $t$, a causal relation between $p$ and $i$ is imposed.

The repayment must be firmly linked to the original transfer of the
principal sum $p$.
How that link is formalized depends on the formalism.
For instance, one may employ a system wide unique indexing of events by
means of a progressive sequence number which can be used to refer to
preceding events.

A savings account with interest is a financial product which is
considered entirely unproblematic and even virtuous in Western finance.
That judgement is not shared by Islamic jurists who claim the presence
of prohibited \emph{riba}.%
\footnote
{We will use \emph{riba} to stand for interest, though this
 identification is by no means unchallenged.
}

\subsection{Justification of the Interest Presence Claim}
\label{subsect-justification-interest}

One may criticize the example for being marginal in terms of interest
payment.
Indeed it seems to incorporate few of the aspects which have
historically led to a condemnation and subsequent prohibition of the
payment of interests.
The importance of savings accounts with interest lies in the role that
they fulfil in Western finance whereas they demonstrably contradict the
prohibition of \emph{riba}.
The justification of the example is as follows:
\begin{itemize}
\item
The risk that $X$ will not be repaid is minimal.
Stronger guarantees of repayment cannot be imagined.
It is likely that the probability of restitution failure is lower than
the probability of theft or loss if the sum is kept by $X$ in his own
physical possession.
\item
The reward $-c + i$ splits into two parts.
The transaction costs $c$ are constant and non-objectionable.
The increment $i = q \cdot p$ is obtained by $X$ without performing any
additional work or investment, or taking any noticeable risks.
Because $-c + i > 0$, the increment $i$ is significant.
\item
$X$ acts out of free will, $X$ is aware of all details of the entire
transaction, $X$ incurs no significant downside risk, $X$ may freely
choose a service provider from a number of offerings mediated by $Z$.
Thus, no other problems exist with the transaction.
\item
It may be concluded that the transaction is unproblematic except for its
incorporation of the increment $i = q \cdot p$.
The qualification of that increment as interest is justified on the
basis of the preceding three points and the fact that a causal relation
between $p$ and $i$ is imposed (see Section~\ref{subsect-details-sawi}).
\end{itemize}

\subsection{Is the Example Pivotal?}
\label{subsect-genericity}

We contemplate the virtue of a systematic investigation of possible
reconstructions of a savings account with interest by synthesis from a
collection of more primitive permissible financial products.
The example of a savings account with interest is pivotal in the sense
that it seems to be the simplest conceivable example that admits a
complete and concise description.
At the same time it is quite realistic.
However, if Islamic jurists would consider a savings account with
interest \emph{halal} after due consideration, the merits of disproving
that it can be synthesized from a collection of other and more primitive
\emph{halal} products disappear, at least from a point of view of
financial system design.
Assuming, however, that all forms of interest are \emph{haram}, a
viewpoint that can be found in many sources, the relevance of a savings
account with interest for our objectives becomes very significant in
view of its  simplicity in comparison to other products which provide
credit for a weaker party in need of financial support.

\section{Preparatory Exploration}
\label{sect-exploration}

Taking a savings account with interest as the central example of a
prohibited financial product, one faces the questions
(i)~how to design a portfolio of permissible basic products,
(ii)~how to specify which compositions of products are permissible,
(iii)~how to prove that a savings account with interest cannot be
synthesized by means of permissible methods of composition from
permissible basic products.

At this point, it needs to be asked to what extent a consistent design
of a reduced product set financial system must comply with the
requirement that a savings account with interest cannot be synthesized
from its basic products.
Intuitively, this seems to be a compelling requirement.
And the fact that provable non-synthesizability for a specific target
product is a meaningful design objective definitely suggests that
techniques from formal methods in computing may be needed to analyze
compliance with that objective.

In the remainder of this section, we carry out a preparatory exploration
of the possibility to synthesize a savings account with interest in the
way known as \emph{tawarruq}.

\subsection{On Interest Free Reduced Product Set Finance}
\label{subsect-if-rpsf}

If a savings account with interest can be synthesized from basic
products in some financial system, this implies that, in functional
terms, what a savings account with interest provides cannot be
forbidden.
Rather it is the way in which a savings account with interest is
described which is disapproved.
As it stands the issue is open as far as we know, and it may be
formulated as a technical question as follows:
\begin{quotation}
\noindent
Are designs of reduced product set financial systems which guarantee
provable non-synthesizability for a savings account with interest
possible, or is one facing the situation that the functionality of a
savings account with interest must be considered unproblematic, because
its synthesis from basic products cannot be prevented in a plausible
way, whereas the particular form of the description of a savings account
with interest is considered problematic?
\end{quotation}
The two possibilities mentioned above are very different and the
understanding of the principles of prohibition of interest depends
critically on resolving this issue.
From the point of view of development of theory of financial systems, it
seems preferable if one or more designs of reduced product set financial
systems can be established which guarantee provable non-synthesizability
for a savings account with interest --- viewed as a functionality.
If that is not possible, an interpretation where the description of a
savings account with interest is simply used as a notational shorthand
for an expanded and functionally equivalent synthesized product can
hardly be avoided.

\subsection{The Progression Architecture to Comply with}
\label{subsect-progression-arch}

In order to prove anything about the possibility to synthesize a savings
account with interest from more primitive products, a limitation of
those primitive products is needed.
The more general form of an impossibility theorem one aims at, the
harder it becomes to map out the preliminaries and to state and prove
the result.
The simplest way to go ahead is to consider known ways to synthesize a
savings account with interest from other primitive products.
In the beginning of the paper, we have sketched three such scenarios:
\emph{tawarruq}, contractus trinus and \emph{murabaha}.
Of these \emph{tawarruq} and contractus trinus synthesize a purely
financial product, whereas  \emph{murabaha} is a sales transaction
involving a non-monetary good.
Contractus trinus involves an insurance policy, which adds complications
that we prefer to avoid.
This leaves us with \emph{tawarruq} as a most relevant scenario at this
stage.
A synthesis of a savings account with interest by means of
\emph{tawarruq} involves the following progression $\pi$ of actions when
put into effect:
\begin{enumerate}
\item
$X$ asks $Z$ to prepare a portion $G$ of a certain good that has exactly
the price $p$;
\item
$X$ buys $G$ at price $p$ from $Z$;
\item
$X$ sells $G$ at price $p - c + i$ to $Y$ where it is agreed that
payment is due at time ${\tt now} + t$;
\item
$Y$ sells $G$ at price $p - \frac{1}{2}c$ to $Z$.
\end{enumerate}
At this level of abstraction, the progression $\pi$ cannot itself be
considered to constitute a synthesis of a savings account with interest
from permissible primitive products.
Rather it constitutes the externally visible behaviour of such a
synthesis, provided the synthesis is successful and also under a number
of conditions that will be uncovered only when attempting to synthesize
a savings account with interest.

We say that $\pi$ is a \emph{progression architecture} for the
synthesized product under investigation.
It represents the requirement that at some level of abstraction the
given actions take place in the given order when putting the synthesized
product into effect.
Whether such a product actually exists is immaterial at this stage.

The plan is to take the progression architecture $\pi$ as a point of
departure and to study the ways to synthesize a savings account with
interest that comply with $\pi$ or a plausible variation of $\pi$.
Putting a product into effect may involve many intermediate steps.

Regarding the the progression architecture $\pi$, the following
assumptions are made:
\begin{itemize}
\item
It is certain that the value of $G$ does not significantly degrade in
the time span $t$.
It is also certain that the value of $G$ does not change in the much
shorter time spans between the four actions of $\pi$.
If $G$ is a portion of a precious metal, it may have this property; but
if it is a portion of a non-durable good, it will not.
\item
$Z$ is able to prepare a portion $G$ of the good at the countervalue of
a sum of money $p$ (within a range that may be implicit in the original
and subsequent detailed descriptions of a savings account with
interest).
Here it may be of importance that $X$ has been able to prepare a sum of
money $p$ that consists of coins and banknotes, which restricts the
range of values that $p$ may take.
\item
The good itself is not money.
Otherwise interest is involved in the third step of the progression
architecture $\pi$.
This means that a philosophical position concerning what constitutes
money is needed (see for instance~\cite{Ber10a} for an attempt to survey
this matter).
Anyhow, the good cannot be used as a measure of value and the values
that a portion can take are less fine grained than monetary values.
For instance, $G$ may be delivered in fixed size blocks of gold and $Z$
may refuse to partition the blocks.
\end{itemize}

\subsection{Refinements of the Progression Architecture}
\label{subsect-refinements}

Because $X$ intends to save, it may be assumed that more money is
available to $X$ and that $X$ can buy a more expensive portion of the
good when needed.
This is relevant in the light of the assumption that the values that a
portion of the good can take are less fine grained than monetary values.
It leads to a variation $\pi'$ of $\pi$ which takes care of the fact
that $X$ is not exchanging money for money with $Z$:
\begin{enumerate}
\item
$X$ requests $Z$ to prepare a portion $G$ of a certain good that has a
price $p'$, where $p < p'$ and $p'$ is chosen by $Z$ as low as possible
such as to be able to deliver $G$ within cost $\frac{1}{2}c$;
\item
$X$ buys $G$ at price $p'$ from $Z$;
\item
$X$ sells $G$ at price $p' - c + i$ to $Y$ where it is agreed that
payment of $p' - p$ is due at once, i.e.~${\tt now}$, and payment of
$p - c + i$ is due at time ${\tt now} + t$;
\item
$Y$ sells $G$ at price $p' - \frac{1}{2}c$ to $Z$.
\end{enumerate}

The real complication that must be dealt with lies in the flow of
contracts that must underly this transaction.
Contracts provide participants with assurances that guarantee that their
actions make sense or, in other words, that their actions will indeed
emerge in the progression that results from multi-threading the
progressions of their individual plans.
In particular:
\begin{itemize}
\item
$Z$ needs to be sure that $X$ will buy $G$ at price $p'$ before he
prepares $G$;
\item
$X$ needs to be sure that $Y$ will buy $G$ at price $p' - c + i$ by
means of a credit transaction before he buys $G$ from $Z$;
\item
$Y$ needs to be sure that $Z$ will buy $G$ back at price
$p' - \frac{1}{2}c$ by means of a spot transaction.
\end{itemize}
At this stage the progression architecture $\pi'$ must be refined in
order to take into account the signing of contracts.
At least three contracts are needed and the contracts involve
conditional obligations.
This leads to $\pi^{\prime\prime}$:
\begin{enumerate}
\item
$X$ and $Z$ sign a contract $C_1$ (where $Z$ takes the initiative) that,
if $Z$ prepares a portion $G$ of a certain good as specified in $\pi'$,
$X$ will subsequently buy it from $Z$ at the price specified in $\pi'$;
\item
$X$ and $Y$ sign a contract $C_2$ (where $X$ takes the initiative), with
reference to contract $C_1$, that, after $X$ has bought $G$ from $Z$,
$Y$ will buy it from $X$ by means of a credit transaction as specified
in $\pi'$;
\item
$Y$ and $Z$ sign a contract $C_3$ (where $Y$ takes the initiative), with
reference to contract $C_2$, that, after $Y$ has bought $G$ from $X$,
$Z$ will buy it from $Y$ by means of a spot transaction as specified in
$\pi'$ (under the assumption that earlier $X$ has bought $G$ from $Z$ in
accordance with contract $C_1$);
\item
$X$ requests $Z$ to prepare a portion $G$ of a certain good that has a
price $p'$, where $p < p'$ and $p'$ is chosen by $Z$ as low as possible
such as to be able to deliver $G$ within cost $\frac{1}{2}c$;
\item
$Z$ informs $X$, with reference to contract $C_1$, that he has prepared
$G$;
\item
$X$ buys $G$ at price $p'$ from $Z$;
\item
$X$ informs $Y$, with reference to contract $C_2$, that he has bought
$G$;
\item
$X$ sells $G$ at price $p' - c + i$ to $Y$ where it is agreed that
payment of $p' - p$ is due at once, i.e.~${\tt now}$, and payment of
$p - c + i$ is due at time ${\tt now} + t$;
\item
$Y$ informs $Z$, with reference to contract $C_3$, that he has bought
$G$;
\item
$Y$ sells $G$ at price $p' - \frac{1}{2}c$ to $Z$.
\end{enumerate}

Having progression architecture $\pi^{\prime\prime}$ for a synthesis of
a savings account with interest at hand, we may start looking for
positive as well as negative results concerning the possibility to
synthesize a savings account with interest.
Positive results depend on conditions concerning the readiness of $Z$
and $Y$ to sign contracts $C_1$ and $C_2$, respectively, when requested
by $X$ to do so.
Readiness requires that all contracts are prepared before any of them is
signed and that they are signed in a plausible order.
Covering the preparation of the contracts and providing a plausible
order in which the contracts must be signed leads to
$\pi^{\prime\prime\prime}$:
\begin{enumerate}
\item
$X$ and $Z$ prepare a contract $C_1$ (where $Z$ takes the initiative)
that, if $Z$ prepares a portion $G$ of a certain good as specified in
$\pi'$, $X$ will subsequently buy it from $Z$ at the price specified in
$\pi'$;
\item
$X$ and $Y$ prepare a contract $C_2$ (where $X$ takes the initiative),
with reference to contract $C_1$, that, after $X$ has bought $G$ from
$Z$, $Y$ will buy it from $X$ by means of a credit transaction as
specified in $\pi'$;
\item
$Y$ and $Z$ prepare a contract $C_3$ (where $Y$ takes the initiative),
with reference to contract $C_2$, that, after $Y$ has bought $G$ from
$X$, $Z$ will buy it from $Y$ by means of a spot transaction as
specified in $\pi'$ (under the assumption that earlier $X$ has bought
$G$ from $Z$ in accordance with contract $C_1$);
\item
$Y$ and $Z$ sign contract $C_3$;
\item
$X$ and $Y$ sign a contract $C_2$;
\item
$X$ and $Z$ sign a contract $C_1$;
\item
The remaining steps of $\pi^{\prime\prime\prime}$ are steps 4 to 10 of
$\pi^{\prime\prime}$.
\end{enumerate}

\subsection{Plan of Technical Work}
\label{subsect-plan}

On the basis of the presented preparatory exploration, it should be
possible to derive formal results concerning the existence and
non-existence of synthesized reconstructions of a savings account with
interest.
General conditions will determine about the matters that have to do with
willingness of involved parties to sign contracts, certainty that signed
contracts are obeyed, fluctuations of the value of the good during
different steps of a progression, knowledge about such fluctuations to
parties involved, and the legality of ignoring such knowledge.
In order to make progress from this stage onwards, attention must be
paid to the development of a portfolio of permissible financial products
at least involving a spot sale, preparing and signing certain contracts,
and informing certain parties.

It seems plausible that at least one credit sale transaction is needed
for any synthesis of a savings account with interest.
However, proving this still constitutes a challenge because it requires
a complete and plausible set of assumptions on available products and
compositions thereof.

A positive result is likely to be found if packaging of the contracts in
a single contract with three signatures is permissible.
This is a deviation of $\pi^{\prime\prime}$ which removes the need to be
able to cope with the preparation of contracts.
Still there is a non-trivial problem about dealing with the consequences
of fluctuations of the value of $G$: $X$ can only expect that an overall
contract $C$ is signed by $Y$ and $Z$ if the fluctuations are considered
unproblematic by them.
If fluctuations may become large, there emerges a risk (\emph{gharar})
which $Y$ and $Z$ may not be entitled to ignore at their own
responsibility.

Another complication with the single contract case is connected with the
status of the good of which $G$ is a portion.
If the value of $G$ is completely stable, which is very helpful for
getting the contract(s) signed, then the shaping of the good into units
of a fixed form and weight that have a higher value than a unit of money
may not suffice to demonstrate that the good itself is not money.
A theory of non-monetary goods is needed in this case to substantiate
the non-circularity of the synthesis of a savings account with interest.
A thorough analysis of the single contract case must be performed before
any significant progress on the multiple contract case can be made.

Of course, one may simply rule out the single contract case by
disallowing it in an ad hoc fashion because it seems too close to a
savings account with interest.
However, in that case the intriguing fact may be missed out that
deriving the prohibition of a single contract deal from a prohibition of
a savings account with interest depends on implicit assumptions about
the possibility to use the good involved as a measure of value and the
fluctuations of the value of portions of that good.
Those assumptions should be made explicit before embarking on
introducing a rather unsystematic additional prohibition.

\section{Concluding Remarks}
\label{sect-conclusions}

We have made an extensive preparation to an investigation of issues
concerning the form of finance that the avoidance of interest gives rise
to in Islamic finance.
This form of finance is the classical example of the kind of finance
where only financial products that are synthesized from a few basic
financial products according to certain principles are considered.
We have coined the term ``reduced product set finance'' for this kind of
finance.

Firstly, we have made an effort to answer the general questions that, in
our opinion, should be answered before the investigation is conducted
from a non-Islamic perspective.
Secondly, we have listed more specific questions about classical
synthesized financial product examples and the prevailing positions
concerning their legality that need answers in order to gain a clear
understanding of synthesized financial products.
Thirdly, we have done a preparatory exploration of the synthesis of a
savings account with interest in the way known as \emph{tawarruq} to
form a reasonable preliminary picture.

We conclude that the investigation is feasible, but that its outcome is
entirely open.
It is plausible that techniques from formal methods in computing can be
applied successfully.
We claim that in the long run results may be fed back to computer
science, which may constitute a justification of the work in itself.

The merit of this paper may be doubted in the absence of solutions to
the main problems raises by it.
Computer scientists often produce solutions to problems for which the
relevance must be analysed or even discovered afterwards.
However, there is the principle of though that, in some cases, it is not
worth to solve a problem if providing a solution of the problem is
needed as a justification to state it.
This princlple is nowadays seldom applied in computer science.
In the case of the problems that this paper is concerned with, we have
made the decision that solving the problems is only worthwhile if they
can be stated convincingly in advance.
The methodological difficulties involved in solving the problems must be
analysed and contemplated beforehand.
Their significance should not be made dependent on their answers.
Ideally, the problems should be stated such that the task of stating the
problems and the task of providing answers may be performed by different
people or teams.

\subsection*{Acknowledgements}
This work has been carried out in the framework of the project
``Thread Algebra for Strategic Interleaving'', which is funded by the
Netherlands Organisation for Scientific Research (NWO).
We thank Farhad Arbab (Leiden University) for a discussion about
\emph{riba} and interest.

\bibliographystyle{splncs03}
\bibliography{IF}

\begin{thebibliography}{10}
\providecommand{\url}[1]{\texttt{#1}}
\providecommand{\urlprefix}{URL }

\bibitem{AlM02a}
Al-Masri, R.Y.: The binding unilateral promise (\textit{wa'd}) in {Islamic}
  banking operations: Is it permissible for a unilateral promise
  (\textit{wa'd}) to be binding as an alternative to a proscribed contract?
  Islamic Economics  15(1),  29--33 (2002)

\bibitem{BBR10a}
Baeten, J.C.M., Basten, T., Reniers, M.A.: Process Algebra: Equational Theories
  of Communicating Processes, Cambridge Tracts in Theoretical Computer Science,
  vol.~50. Cambridge University Press, Cambridge, UK (2010)

\bibitem{Bak06a}
Bakhshi, A.M.: Developing a Financial Model for {Islamic} Credit Card for the
  {UK}. Master's thesis, University of Salford, Salford, UK (2006)

\bibitem{BDW10a}
Bassens, D., Derudder, B., Witlox, F.: Searching for the {Mecca} of finance:
  {Islamic} financial services and the world city network. Area  42(1),  35--46
  (2010)

\bibitem{Ber10a}
Bergstra, J.A.: Formaleuros, formalcoins and virtual monies. {\tt
  arXiv:1008.0616 [cs.CY]} (August 2010)

\bibitem{Ber10b}
Bergstra, J.A.: Informal control code logic. {\tt arXiv:1009.2902 [cs.SE]}
  (September 2010)

\bibitem{BL02a}
Bergstra, J.A., Loots, M.E.: Program algebra for sequential code. Journal of
  Logic and Algebraic Programming  51(2),  125--156 (2002)

\bibitem{BM04c}
Bergstra, J.A., Middelburg, C.A.: Thread algebra for strategic interleaving.
  Formal Aspects of Computing  19(4),  445--474 (2007)

\bibitem{BM07b}
Bergstra, J.A., Middelburg, C.A.: Machine structure oriented control code
  logic. Acta Informatica  46(5),  375--401 (2009)

\bibitem{BM09a}
Bergstra, J.A., Middelburg, C.A.: Timed tuplix calculus and the {Wesseling} and
  van den {Bergh} equation. {\tt arXiv:0901.3003 [q-fin.GN]} (January 2009)

\bibitem{BP09b}
Bergstra, J.A., Ponse, A.: A progression ring for interfaces of instruction
  sequences, threads and services. {\tt arXiv:0909.2839 [cs.PL]} (September
  2009)

\bibitem{BT07a}
Bergstra, J.A., Tucker, J.V.: The rational numbers as an abstract data type.
  Journal of the ACM  54(2),  Article 7 (2007)

\bibitem{BR08a}
Bethke, I., Rodenburg, P.H.: The initial meadows. {\tt arXiv:0806.2256
  [math.RA]} (June 2008)

\bibitem{Bur07a}
Burgess, M.: System administration and the scientific method. In: Bergstra,
  J.A., Burgess, M. (eds.) Handbook of Network and System Administration, pp.
  689--728. Elsevier, Amsterdam (2007)

\bibitem{Cha85a}
Chapra, M.U.: Towards a Just Monetary System, chap.~5. The Islamic Foundation,
  Leicester (1985)

\bibitem{CL09a}
Chong, B.S., Liu, M.H.: {Islamic} banking: Interest-free or interest-based?
  Pacific-Basin Finance Journal  17(1),  125--144 (2009)

\bibitem{ElG01a}
El-Gamal, M.A.: An economic explication of the prohibition of \textit{gharar}
  in classical {Islamic} jurisprudence. Islamic Economic Studies  8(2),  29--58
  (2001)

\bibitem{ElG06a}
El-Gamal, M.A.: {Islamic} Finance: Law, Economics and Practice. Cambridge
  University Press, Cambridge, UK (2006)

\bibitem{ElG08a}
El-Gamal, M.A.: Incoherence of contract-based {Islamic} financial jurisprudence
  in the age of financial engineering. Wisconsin International Law Journal
  25(4),  605--623 (2008)

\bibitem{GW09a}
Gait, A.H., Worthington, A.C.: A primer on {Islamic} finance: Definitions,
  sources, principles and methods. Discussion Papers Finance 2009-09,
  Department of Accounting, Finance and Economics, Griffith University (2009)

\bibitem{HR04a}
Huth, M., Ryan, M.: Logic in Computer Science. Cambridge University Press,
  Cambridge, UK (2004)

\bibitem{Iqb10a}
Iqbal, M.M.: Prohibition of interest and economic rationality. Arab Law
  Quarterly  24(3),  293--308 (2010)

\bibitem{Jan08a}
Janlert, L.E.: Dark programming and the case for the rationality of programs.
  Journal of Applied Logic  6(4),  545--552 (2008)

\bibitem{Kam09a}
Kamla, R.: Critical insights into contemporary {Islamic} accounting. Critical
  Perspectives on Accounting  20(8),  921--932 (2009)

\bibitem{KM08a}
Kr{\"{o}}ger, F., Merz, S.: Temporal Logic and State Systems. Texts in
  Theoretical Computer Science, An EATCS Series, Springer-Verlag, Berlin (2008)

\bibitem{Lew99a}
Lewison, M.: Conflicts of interest? {The} ethics of usury. Journal of Business
  Ethics  22(4),  327--339 (1999)

\bibitem{Mid10a}
Middelburg, C.A.: Searching publications on operating systems. {\tt
  arXiv:1003.5525 [cs.OS]} (March 2010)

\bibitem{Mid10b}
Middelburg, C.A.: Searching publications on software testing. {\tt
  arXiv:1008.2647 [cs.SE]} (August 2010)

\bibitem{NA09a}
Noor, A.M., Azli, R.M.: A review of {Shariah} compliant instruments for
  {Islamic} credit cards as adopted by {Malaysian} financial institutions.
  International Journal of Monetary Economics and Finance  2(3--4),  221--238
  (2009)

\bibitem{PZ06a}
Ponse, A., van~der Zwaag, M.B.: An introduction to program and thread algebra.
  In: Beckmann, A., et~al. (eds.) CiE 2006. Lecture Notes in Computer Science,
  vol. 3988, pp. 445--458. Springer-Verlag (2006)

\bibitem{Sal06a}
Salvatore, A.: Power and authority within {European} secularity: From the
  enlightenment critique of religion to the contemporary presence of {Islam}.
  The Muslim World  96(4),  543--561 (2006)

\bibitem{Sid02a}
Siddiqi, M.N.: Comparative advantages of {Islamic} banking and finance. In:
  Proceedings of the Fifth Harvard University Forum on Islamic Finance. Harvard
  University (2003)

\bibitem{Wol05a}
Wolters, W.: Is een islamitische economie mogelijk? Valedictory lecture,
  Radboud Universiteit Nijmegen, ISBN 90-9019111-9 (2005)

\end{thebibliography}


\end{document}